\documentstyle[12pt,epsf,fleqn]{article}
\renewcommand{\baselinestretch}{1.3}

\begin{document}
\title
{\bf A XENON SOLAR NEUTRINO DETECTOR}

\author
{A.Sh.~Georgadze$^*$, H.V.~Klapdor-Kleingrothaus$^\dagger$, H. P\"as$^\dagger$
\\
and Yu.G.~Zdesenko$^*$\\
$^{*}${\it Institute for Nuclear Research, 252028, Kiev, Ukraine}\\
$^{\dagger}$ {\it Max-Planck-Institut f\"ur Kernphysik,}\\ 
{\it
P.O. Box 103980,
D-69029 Heidelberg, Germany}}
\date{}
\maketitle

\begin{abstract}
\renewcommand{\baselinestretch}{1.0}
\small\normalsize
\noindent
The
neutrino capture 
by $^{131}$Xe with a threshold at 352 keV as reaction
to detect solar 
neutrinos is examined. The most important feature of
this
process is its high sensitivity to beryllium neutrinos, which
contribute
approximately 40\% to the total capture rate (45 SNU) assuming the fluxes
predicted by
the Standard
Solar Model of \cite{SSM1}. 
The procedure of extraction of the daughter
cesium 
atoms from liquid xenon 
as well as other technical problems
concerning preparation of the cesium 
sample,
low background measurements
and side reactions for a possible realisation as
a solar neutrino detector
are discussed. 
The expected counting rate according to the SSM of \cite{SSM1}
for a $^{131}$Xe
detector 
is $\approx$ 1500 events/y$\cdot$kton. The results of
such a detector will be important for solving the puzzle of
the possible existence  
of vacuum oscillations or the MSW effect and/or input
parameters
of the Standard Solar Models.
\end{abstract}

\thispagestyle{empty}
\renewcommand{\baselinestretch}{1.3}
\small\normalsize
\section{Introduction}

Reconciling the Kamiokande and Homestake experiments
one was led to the the conclusion 
that not only $^{8}$B but $^{7}$Be
neutrinos suffer considerable
reduction with respect to predictions
of the SSM. 
The GALLEX \cite{GALLEX}
and SAGE \cite{SAGE} results support this conclusion. In particular, the SSM
prediction for the gallium experiment is 113 -- 132 SNU
\cite{SSM1},\cite{SSM2},\cite{Dar}.
Since the pp neutrino flux is
almost
model independent and closely tied with the solar luminosity it 
was
supposed that its contribution to the GALLEX result can be predicted with
reasonable precision to be 70 SNU \cite{SSM1}. The boron neutrino
contribution fixed by the Kamiokande data
in model independent way is equal
to 7 SNU. 
Thus, comparing theoretical and experimental results
one can see
that in the gallium data there is no room for beryllium
neutrinos which are
expected to contribute 31 -- 34 SNU or 25 --
27\% of the total flux in the
SSMs. 
After calibration of the GALLEX experiment with a
$^{51}$Cr neutrino
source a
deficit of $^{7}$Be neutrinos was claimed at a 3$\sigma$
confidence level \cite{Kirst}.

Efforts to explain the reduction of
beryllium neutrinos on
the basis of variation of the input parameters of the standard solar
models were unsuccessful
until now according to \cite{Bahc}.

Therefore the $^{7}$Be neutrino deficit should be proved in another 
experiment with essential sensitivity to beryllium neutrinos.

In fact, the most important parameter
of the detector is the 
relative ratio of the  $^{7}$Be to pp
neutrino
yield, which should be larger than in the $^{71}$Ga experiments. 
The
latter means
that the threshold of the detector should be low enough to
avoid possible effects
of
reduction of the $^{7}$Be neutrino capture cross section near the
threshold, and 
high enough to reduce the relative contribution of the pp
neutrinos in the total yield.

There is a number of solar neutrino
detectors proposed for the near and far 
future. Two of them,
Superkamiokande \cite{Suz2} (in operation now) 
and SNO \cite{SNO} (Sudbury
Neutrino Observatory which
will start data taking in the nearest future)
are sensitive to boron neutrinos mainly.
Therefore these experiments will
give no full solution of the problem. 
The same conclusion is true
for
ICARUS  \cite{ICARUS} (a multipurpose liquid argon drift 
detector which
will operate in the 
Gran Sasso Underground Laboratory). 
The HELLAZ
detector \cite{HELLAZ} (2000 m$^{3}$ helium TPC at 5 bar, 77K) would
be
sensitive to pp and $^{7}$Be 
neutrinos but is far away from real
operation, as many other proposals like HERON 
 \cite{HERON}, $^{81}$Br
\cite{81Br}, $^{115}$In  \cite{115In} and detectors
based on low
temperature techniques   \cite{Fior1}, because all of them are in 
very
early stages of their development. 

Only two of the proposals, BOREXINO
and an iodine  
detector, that both would be able to detect $^{7}$Be
neutrinos should be on
line relatively soon.

The main goal of the BOREXINO
 \cite{BOREXINO} experiment is to measure
 the flux of $^{7}$Be 
neutrinos
via elastic scattering in a 100 tons liquid scintillator target. 
Within
the electron energy window $0.25<E_{e}<0.80$ MeV 80\% of the signal
comes
from $^{7}$Be neutrinos leading to a high signal rate of 50 counts
per day. It
is supposed to use an ultra-pure water, liquid scintillator and
selected constructing materials 
to reduce the background of the detector
to the extremely low level of 0.1 
counts/day$\cdot$ton (for a signal/background
ratio of 5:1). 
To date a prototype of the BOREXINO 
detector with a volume
of scintillator of 4.5 m$^{3}$ was mounted and tested 
successfully in the
Gran Sasso Underground Laboratory \cite{BOREXINO}. 
An
extremely high purity of the scintillator at the level of
 $\approx(2-5)\cdot$10$^{-16}$ g/g for $^{238}$U and
$^{232}$Th contamination was demonstrated. 
The background of the prototype detector in the
energy region
of 250-800 keV was $\approx$ 1800 counts/day \cite{BOREXINO}
which is however still 
much higher than required. Therefore still big efforts and
time will be needed 
to reach the required sensitivity. 
 
An experiment based on the 
$^{127}$I+$\nu \rightarrow
^{127}$Xe$^\ast +e^{-}$
neutrino capture reaction to the 3/2$^{+}$ excited
state of $^{127}$Xe with the
threshold at E$_\nu$=0.789 MeV was suggested by Haxton and is in the pilot
research stage now \cite{127I}.
However this project suffers of the strong model dependent theoretical 
estimation of the total capture
rate, ranging between 20 and 135
SNU \cite{Lut}. Also experimental efforts as the measurement of
the charge-exchange reaction     
$^{127}$I~(p,n)~$^{127}$Xe, which could not be  
normalized to known beta  transitions, as well as the attempt to 
fix the GT distribution in an experiment at LAMPF,
\cite{LAMPF} 
where the involved neutrinos had high 
multipolarities due to their somewhat higher energies, were of limited value
in this respect.

Therefore
it will be difficult to solve the $^{7}$Be neutrino problem within a 
reasonable time scale using present detectors,
and a new detector would 
be desirable. We  
discuss here the
possibility to use $^{131}$Xe as radiochemical detector
 with a threshold at
352 keV \cite{NDS} and with good sensitivity to 
$^{7}$Be neutrinos. 
This possibility
is based first on the excellent 
properties of xenon which allow to use it
as a target in a radiochemical experiment as well as on  
a
reliable prediction of the neutrino capture rate in
$^{131}$Xe \cite{APP}.

\section{Cross sections and capture
rates}

The proposed detection process is based on the neutrino capture
reaction 
$\nu +^{131}$Xe$\rightarrow ^{131}$Cs+$e^{-}$.
The low threshold
of this reaction allows 
to record
approximately one fifth of the solar
neutrino flux from the $p-p$ reaction
and both lines from $^{7}$Be decay.

The reduced transition probability for
neutrino capture by the $^{131}$Xe g.s.
can be derived from electron capture of
$^{131}$Cs. The
half-life of $^{131}$Cs is $T_{1/2}=9.689$ 
days \cite{NDS}
and log($f_{EC}t_{EC}$)=5.53 (the
$f_{EC}$ value is taken from \cite{Gov}),
which gives [$B^{+}_{F}+B^{+}_{GT}$]=0.018.

The cross section for neutrino
capture with energy $E_{\nu}$ is defined
as

$\sigma_{\nu}(E_{\nu})=\frac{g_{A}^2g_{V}^2}{\pi c^3 \hbar^4}
\sum_{E'}
p_e E_e F(Z,A,E_e)
B_{GT}(E')$~~~~~~~~~~~~~~~~~~~~~~~~~~~~~~~~~~~~~~~~~~~~~~~~~~~~~~~~~(1)

with the nuclear excitation energy $E'$, the electron momentum $p_e$ and
energy $E_e$, the Gamow--Teller matrix elements $B_{GT}$ 
and the Fermi
function $F(Z,A,E_e)$ according to \cite{Gov}. 

The capture rates
predicted for $^{131}$Xe by the SSM of \cite{SSM1} for the 
ground-state to ground-state
transition are presented in Table 1 (in raw a).
They indicate the minimum
neutrino capture rate, based on the experimentally 
known transition 
strength.
To take into account the contribution of the excited states 
we have
performed pn--QRPA calculations following the procedure given in 
\cite{B},
\cite{C}. 
The chosen parameter set included the Nilsson 
potential parameter
of Ragnarsson and Sheline
\cite{3}, the pairing interaction strength fitted
to reproduce the energy
gap, the nuclear deformation obtained from the mass 
formulae of Moeller
and Nix \cite{4} for the daughter nucleus
$^{131}$Cs and the strength of
the Gamow Teller interaction 
obtained by a fit of experimental half life
values of the isotope series of the
daughter nucleus Cs. This parameter set
gave the best reproduction
of experimental data.  The ground state matrix
element 
0.018 was taken from the experimental half-life of $^{131}$Cs. The
proton 
separation threshold of 5.47 MeV excitation energy in $^{131}$Cs
was taken from
\cite{waps}.

From the level scheme \cite{Led} one can see that
four excited states (except the forbidden
transition to the $7/2^{+}$ level)
are below the $^{7}$Be neutrino
capture
threshold. Neutrino capture to these states will increase the part
of the 
beryllium neutrinos in the total absorption rate compared 
to the
ground state capture.

The Gamow-Teller strength calculated by pn--QRPA
results for beryllium 
neutrinos in a capture rate of 17.8 SNU which is only slightly
larger than capture to the ground-state of
$^{131}$Cs alone (13.6 SNU, see Table 1).  
However, the determination of matrix
elements for the transitions to the low-lying excited 
states, as needed
for $pep$, $^{7}$Be, $^{13}$N, $^{15}$O neutrinos, in nuclear structure
calculations, 
includes considerable uncertainties.

To check the 
parameters of our model calculations
we have evaluated the contribution of
these states phenomenologically
from the systematics of known beta-decay
rates in the vicinity of the mass
number 131. 
There are some $\beta^{-}$
decays of $^{127}$Te$\rightarrow^{127}$I,
$^{129}$Te$\rightarrow ^{129}$I,
$^{133}$Xe$\rightarrow ^{133}$Cs, $ ^{135}$Xe$\rightarrow^{135}$Cs, which
are 
similar to the $^{131}$Xe$\rightarrow ^{131}$Cs
$(3/2^{+})_{\hbox{g.s.}}\rightarrow(5/2^{+})$ transition.

For all these
cases the transition strength to the first $5/2^{+}_{1}$ state is 
systematically several times larger then
for the $5/2^{+}_{2}$ state. 
The $3/2^{+}\rightarrow 3/2^{+}$ and
$3/2^{+}\rightarrow 1/2^{+}$
transitions are considerably weaker compared
to the transition 
to the 5/2$^{+}$ excited level.
Consequently, according to these arguments the
contribution of the low-lying excited states to the beryllium
rate
estimated in this way would not exceed 5\% of the ground state rate.

The cross section for boron neutrinos cannot be inferred reliably
from
beta-decay systematics due to the large increase in
the capture cross section
for $^{8}$B neutrinos caused by the GT
strength to exited states with
energies between 2-7 MeV above the
ground state of $^{131}$Cs. The QRPA
calculations produce for boron 
neutrinos a rate of 12.7 SNU which is
stable enough to variation
of model parameters to an accuracy of
20--30\%.
The obtained capture rate for boron neutrinos 
is comparable with
the contribution of boron neutrinos for other heavy nuclei,
$^{71}$Ga,
$^{81}$Br, $^{127}$I, $^{115}$In (see Table 1).  

We have calculated a
total neutrino capture rate for $^{131}$Xe  of 
45.2 $^{+12}_{-6}$ SNU assuming the fluxes predicted by the SSM of
\cite{SSM1}.
The uncertainty of the total
capture rate 
is estimated to be of the order of 25--30\% and has been
determined mainly by variations of the parameters in the model calculations
for the transitions to the highly exited states of $^{131}$Cs. The possible
(p,n) measurement of the transition strength to excited states
may allow
to reduce the uncertainty of the total capture rate. 
It
should be mentioned also that the xenon detector can be calibrated with a
$^{51}$Cr 
neutrino source, 
as it was successfully performed for  GALLEX
\cite{GALLEX}. 
Such a procedure can precisely 
determine the efficiency of
beryllium neutrino detection and 
will decrease the uncertainty 
of the
contribution of the excited states.
All described points lead to  the
statement that the  
uncertainties for the proposed 
xenon detector are
clearly smaller then for other detectors suggested to 
measure beryllium neutrinos.

The predicted integral rate of
absorption events (including the contribution
of the excited states) for the
xenon detector from the main neutrino sources
is, for the SSM of \cite{SSM1}, $R \simeq 1500$
events$/kton\cdot y$. 

\section{Remarks on a possible detector realisation
and the background}

The half-life of $^{131}$Cs ($t_{1/2}= 9.688  d)$
and the physical
properties of xenon allow to build up a detector 
of
neutrino capture events based on the extraction of the reaction 
products
$^{131}$Cs from the xenon target and subsequent  
observation of their
decays.

The abundance of $^{131}$Xe is as large as 21.18\% and therefore
natural xenon
may be used as a target. 
For reasonable dimensions of the
vessel the xenon must be kept in liquid form.
The density of liquid xenon is
3.06 g/cm$^{3}$, the boiling temperature
(165 K) is higher than for
nitrogen, thus it is not difficult from a technical 
point of view to build
up a liquid xenon target with a mass of one kiloton. 
Note that a dewar for
600 tons of liquid argon (i.e. $\approx$ 1.4 kton of liquid xenon)
was designed
for the first stage of the ICARUS experiment \cite{ICARUS}. 

For cesium
extraction the considerable
difference of chemical properties of xenon and
cesium can be exploited. The ionization potential of cesium atoms (3.8 eV) is
the 
lowest among all elements.
The Cs$^{+}$ ions formed in the process of
neutrino capture will be kept
in this state because the ionization
potential  of xenon atoms is much 
 higher (12.1 eV), therefore there will
be no free electrons to recombine with
cesium ions. Electrons can be
produced by natural radioactivities
and cosmic rays but estimations show
that such  processes will be 
negligible if the detector will be placed in
a proper underground environment
and the purity of xenon will be kept on
a sufficient level.  Xenon is an inert gas which can be purified to the
level of better than 1 ppb
(residual oxygen less than 0.1 ppb)
\cite{Belli}, \cite{Benetti}.
 
Despite of many still unknown details of
the behaviour of
the Cs$^{+}$ ions
in liquid xenon, the possible procedure of the extraction
of the daughters would 
be developed on the base of the methods described
in \cite{Mitch}, \cite{Mijad}, where the behaviour of Ba$^{++}$ ions in
liquid and gaseous xenon was studied (with the aim to detect the daughter
products of double beta decay of $^{136}$Xe).  
In the way as was suggested
in \cite{Mijad} 
one or several electrodes  can be placed in the
dewar with
xenon and biased with high negative potential which supplies an
electric
field of about 1 kV/cm in the vessel volume. The drifting mobility
of positively charged ions in liquid 
xenon is of the order of
3$\cdot$10$^{-4}$ cm$^{2}$/V$\cdot$s 
\cite{Dav}, thus the collection time
of produced Cs$^{+}$ ions over
 several meters distance will not exceed
half an hour. Collected
on the electrode, the Cs$^{+}$ ions will be frozen
on the surface
and kept there for the whole exposure time. Then the
collector 
electrode will be removed from the dewar (with proper
precautions to exclude
any possible contamination) to prepare the cesium
sample, following the technique which was developed  in \cite{Nei}.

The
measurement of the electron capture of $^{131}$Cs to $^{131}$Xe
can be
performed by means of detection of the Auger electrons and X-rays in a 
low
background counter. Due to the relatively high energies of the $K$- and
$L$-lines in  $^{131}$Cs  decay, which are
35 keV and 5 keV, respectively,
a possible method of measurement could be developed on the base of  Si
semiconductor detectors and a thin solid  $^{131}$Cs sample introduced
between them. 

One of the most crucial issues for the detection of
neutrino
capture events is the background related with side reactions that
may
produce cesium isotopes decaying by electron capture with life
time
long enough in order to penetrate to the cesium
counter. 
There are
four cesium isotopes, $^{129}$Cs, $^{130}$Cs, $^{131}$Cs 
and $^{132}$Cs,
whose half-lives are longer than several seconds. The contribution of
$^{130}$Cs  is limited due
to the low abundance of this isotope of 4.1\%. 
Two of them,  $^{129}$Cs and
$^{132}$Cs are
decaying to exited states of xenon and therefore can be
eliminated by
means of registration of the $\gamma$-rays.  The same is true
for  $^{134}$Cs, $^{136}$Cs and $^{137}$Cs, which may be present in xenon
or can be produced in spontaneous fission of  $^{238}$U. They decay to
excited states of barium producing characteristic gamma-
rays and therefore
can be eliminated.

Thus mainly $^{131}$Cs created in side reactions can
produce a 
substantial background.

Protons appear mainly as products of
cosmic muons. The relevant   
reaction  $^{131}$Xe$(p,n)^{131}$Cs is
usually a main source of background
in solar neutrino experiments. The
cross-section of this 
reaction is presently unknown. The contribution of
this reaction
can be determined in direct measurement of the reaction
products in the way
as it was done in the GALLEX experiment. In this case
some amount of concentrate
solution was exposed to a muon beam at CERN.

Neutrons from
natural radioactivity with energies of several
MeV may take part in
(n,$\gamma )$ reactions. For these reactions there is
no path to $^{131}$Cs
via $^{130}$Xe(n,$\gamma )^{131}$Xe followed by beta decay to
cesium, since
$^{131}$Xe is stable.
Fast neutrons produced by cosmic muons may induce the
reaction
$^{131}$Xe(n,p)$^{131}$I followed by $^{131}$Xe(p,n)$^{131}$Cs but
its yield is small 
compared to reactions with primary cosmic rays protons.

Cesium isotopes can be produced in spontaneous fission of possible
$^{238}$U impurities.
Fortunately the relative yield of
$^{131}$Cs is quite small, since the cesium isotopes 
fission products
appear with mass numbers mainly
above 136. The xenon purification to a
level of 10$^{-15}$ g/g
for uranium and thorium impurities will reduce the
amount of $^{131}$Cs to a 
negligible level.

Alpha particles may take part
in the following reactions producing $^{131}$Cs:
 $^{128}$Xe $(\alpha ,n)^{131}$Ba (followed by electron capture to $^{131}$Cs);
$^{128}$Xe$(\alpha ,p)^{131}$Cs. 
Secondary protons from the latter
reaction also may produce 
cesium via (p,n) reactions.
The naturally
occurring alpha particles from the
uranium and thorium series have average
energies of several MeV with
maximum at 9.0 MeV which is much smaller than
the Coulomb barrier of
17.5 MeV. Therefore cross sections of these
reactions are strongly suppressed.
Alpha particles also appear as
spallation products of cosmic muons 
with energy high enough to drive these
reactions. However the muon 
flux deep underground
is very weak (e.g., less
than $1/(m^{2}\cdot h)$ at the Gran Sasso Laboratory),
and since the cross
sections are very small (on the order of $10^{-29}cm^{2}$ 
per
nucleon) the
higher-energy alpha particles arising in the detector
volume will not
produce more then a few cesium
atoms per year.\\
It should be stressed,
however, that all mentioned points concerning the extraction of  $^{131}$Cs
ions, preparation and measurement of the sample as well as careful
consideration of the background have to be the subject of additional
investigation.

\begin{table}
\renewcommand{\baselinestretch}{1.0}
\small\normalsize
Table 1:
Capture rates predicted by the standard solar model of [3]
when using the Gamow-Teller strength distribution calculated in this work
for $^{131}$Xe, and capture rates for some other targets from [3] , 
given in SNU (a -- capture rates for the ground-state to ground-state transitions; b -- capture rates including transitions to excited states).
\vspace{5mm}

\begin{tabular}{|l|c|c|c|c|c|c|c|}
\hline
Isotope  & p-p  & pep & $^{7}$Be & $^{8}$B  & $^{13}$N &
 $^{15}$O & Total\\
\hline
\hline
$^{37}$Cl & 0.0 & 0.2 & 1.1 & 6.1 & 0.1 & 0.3 & 7.9\\
\hline
$^{40}$Ar & 0.0 & 0.0 & 0.0 & 1.7 & 0.0 & 0.0 & 1.7\\
\hline
$^{71}$Ga & 70.8 & 3.0 & 34.3 & 14.0 & 3.8 & 6.1 & 132.0\\
%$^{81}$Br & 0.0 & 1.1 & 8.6 & 15.3 & 0.9 & 1.9 & 27.8\\
%$^{98}$Mo & 0.0 & 0.0 & 0.0 & 17.3 & 0.0 & 0.0 &  17.4\\
\hline
$^{115}$In & 468.0 & 8.1 & 116.0 & 14.4 & 13.6 & 18.5 & 639.0\\
\hline
$^{127}$I&    & 1.85 & 14.0 & 18.4 & 0.727& 2.43 & 36.4 \\
\hline
~~~~~~~~ a    & 8.7  & 0.9  & 13.6 &  0.5  & 1.1 & 1.3 & 26.1\\
$^{131}$Xe    &      &      &      &       &      &      &     \\
~~~~~~~~ b    & 9.7  & 1.6  & 17.8 & 12.7  & 1.6 & 1.8 & 45.2 \\
\hline
\end{tabular}
\end{table}

\renewcommand{\baselinestretch}{1.3}
\small\normalsize
\section{Conclusion}

A solar neutrino problem evidently
exists ( in particular for  $^{7}$Be 
neutrinos). Several possible
solutions 
including ``terrestrial'', ``astrophysical''
and
``particle physics'' solutions were considered.  
Therefore the $^{7}$Be 
neutrino deficit should be proved by another detector with a
better relative 
ratio of the $^{7}$Be to the pp neutrino detection than
$^{71}$Ga experiments.
We have outlined why in addition to the present efforts in this 
direction, as BOREXINO, a further
detector is needed.  

In the present work a new solar neutrino detector is
proposed which meets
this requirement. It is based on neutrino capture
in
$^{131}$Xe with a threshold at 352 keV. The total capture rate for 
this
reaction is calculated to be equal to 45.2 SNU. 
A one kiloton  detector
will produce a net neutrino signal of
$\simeq$ 1500 events/yr. The main
advantage of the 
proposed xenon detector is the high sensitivity to
$^{7}$Be neutrinos,
which contribute $\simeq$ 37\% of the total capture
rate.

The uncertainties of the calculated capture rates reflect the
contributions
from the highly excited states of $^{131}$Cs and may be
decreased 
significantly by measuring the transition strength to excited
states in
(p,n) reactions and by calibration of the $^{131}$Xe detector
with a $^{51}$Cr
neutrino source. Due to the high abundance of $^{131}$Xe
(21.18\%) and the physical
properties of xenon, there are no 
technical restrictions 
to build up the proposed detector with a mass of one kiloton of 
natural liquid xenon. A procedure of extraction of the daughter
cesium atoms 
from xenon is proposed which is based on collection of
Cs$^{+}$ ions on a 
collector biased with negative potential. The
additional research and development for all technical subjects as well as
for contributions of side reactions are in progress now.

The
feasibility of the proposed xenon detector depends on the
particular cost
of the xenon. Since the price of xenon in the last years fell down
significantly, the cost of large quantities of this gas, and even of enriched
$^{131}$Xe, is no longer
prohibitive. 

The comprehensive analysis of data from the xenon detector
together
with the results of the gallium, chlorine and Superkamiokande
experiments would
allow to determine the $pp,~ ^{7}$Be and $^{8}$B neutrino
fluxes.

This
information will be important to constrain solar parameters, or check the
existence of vacuum oscillations
and MSW effect and finally may help to 
solve one of the most longstanding puzzles
of
modern physics.

\section*{Acknowledgements}
We thank F.A. Danevich and M. Hirsch for useful discussions. 
A.Sh.G. and Yu.G.Z.
are supported by the Science and Technology Center of Ukraine (contract
No. 411). 

\renewcommand{\baselinestretch}{1.0}

\end{document}